\documentclass[10pt]{article}                           

\InputIfFileExists{epsf.sty}{}{}

\begin{document}

\def\today{Re\c cu le 5~d\'ecembre~2002;
accept\'e le 10~f\'evrier~2003}

\def \Oeuvres{O$\!$euvres}
\def \AP{AP}
\def \cad{c'est-\`a-dire}
\def \Lecons{Le\c cons}
\def \D {\hbox{d}}
\def \Log {\mathop{\rm Log}\nolimits}
\def \mod#1{\vert #1 \vert}
\def \sech{\mathop{\rm sech}\nolimits}

\def\jmax{J}

\def\PTP{Prog.~Theor.~Phys.~}



\title{
   Sur la ``solution analytique g\'en\'erale'' d'une \'equation
   diff\'erentielle chaotique du troisi\`eme ordre
\footnote{
IRMA Lectures in  Mathematics and Theoretical Physics {\bf }
(de Gruyter, Berlin, 2003).
Journ\'ees de calcul formel, IRMA, Strasbourg, 21--22 mars 2002.
Correspondance~: RC. S2002/070. nlin.PS/0302056
}
}

\author{T.-L.~Yee\dag, R.~Conte\dag, and M.~Musette\ddag
{}\\
\\ \dag Service de physique de l'\'etat condens\'e, CEA--Saclay
\\ F--91191 Gif-sur-Yvette Cedex, France
\\ Yee@drecam.saclay.cea.fr, Conte@drecam.saclay.cea.fr 
{}\\
\\ \ddag Dienst Theoretische Natuurkunde, Vrije Universiteit Brussel
\\ Pleinlaan 2, B--1050 Brussels, Belgium
\\                     {\rm MMusette@vub.ac.be}
}

\maketitle

{\vglue -10.0 truemm}
{\vskip -10.0 truemm}

\begin{abstract} 
\textit{R\'esum\'e}.
M\^eme si elle est non-int\'egrable, 
une \'equation diff\'erentielle peut n\'eanmoins
admettre des solutions particuli\`eres globalement analytiques.
Sur l'exemple du syst\`eme dynamique de Kuramoto et Sivashinsky,
g\'en\'eriquement chaotique et d'un grand int\'er\^et physique,
nous passons en revue diverses m\'etho\-des,
toutes fond\'ees sur la structure des singularit\'es,
permettant de caract\'eriser la solution analytique qui d\'epend
du plus grand nombre possible de constantes d'int\'egration.

\textit{English abstract}.
Even if it is nonintegrable,
a differential equation may nevertheless admit particular solutions
which are globally analytic.
On the example of the dynamical system of Kuramoto and Sivashinsky,
which is generically chaotic and presents a high physical interest,
we review various methods,
all based on the structure of singularities,
allowing us to characterize the analytic solution which depends
on the largest possible number of constants of integration.
\end{abstract}


\noindent \textit{Mots clefs} :
ondes solitaires,
\'equations de Briot et Bouquet,
Nevanlinna,
algorithme LLL,
\'equation de Kuramoto et Sivashinsky,
\'equation de Ginzburg et Landau complexe cubique unidimensionnelle,
genre,
troncature.

\noindent \textit{PACS 1995}~:
 02.30.-f   
 02.60.-x   
 02.70.-c   
 05.45.+b   
 47.27.-i   
 82.40.-g   

\baselineskip=12truept 


\tableofcontents

\vfill \eject

\section{Situation du probl\`eme}

D'apr\`es un r\'esultat classique \cite{Ruelle},
un syst\`eme diff\'erentiel ne peut pr\'esenter un comportement chaotique
que si son ordre \'egale au moins trois.
D'autre part, malgr\'e l'opposition intuitive entre chaos et analyticit\'e,
il n'est pas interdit \`a un syst\`eme g\'en\'eriquement chaotique 
de poss\'eder des solutions particuli\`eres globalement analytiques,
la seule impossibilit\'e \'etant que le nombre de constantes d'int\'egration
ne peut \'egaler l'ordre.
Le but de cet article est de passer en revue les diverses m\'ethodes
connues permettant de caract\'eriser (id\'ealement d'exhiber)
cette solution analytique, forc\'ement particuli\`ere,
qui d\'epend du nombre maximal possible de constantes d'int\'e\-gration.
Convenons de l'appeler \textit{solution analytique g\'en\'erale},
par opposition \`a la \textit{solution g\'en\'erale} dont l'expression
analytique n'existe pas.

L'exemple physique optimal pour illustrer ces m\'ethodes
est \`a notre avis
l'\'equation de Ginzburg et Landau cubique unidimensionnelle (CGL3)
\begin{eqnarray}
{\hskip -10.0 truemm}
& &
i \psi_t + p \psi_{xx} + q \mod{\psi}^2 \psi - i \gamma A =0,\
p q \gamma \not=0,\
(\psi,p,q) \in {\mathcal C},\
\gamma  \in {\mathcal R},
\label{eqCGL3}
\end{eqnarray}
pour les raisons suivantes,
\begin{enumerate}
\item
elle est g\'en\'eriquement chaotique \cite{MannevilleBook},

\item
elle admet une limite int\'egrable $\Im(p)=\Im(q)=\gamma=0$
(\'equation de Schr\"odinger nonlin\'eaire (NLS)) 
dont tous les \'el\'ements d'\'int\'egrabilit\'e sont connus analytiquement,

\item
les ph\'enom\`enes physiques qu'elle mod\'elise sont tr\`es importants~:
propagation du signal dans une fibre optique \cite{AgrawalBook},
intermittence spatio-temporelle en turbulence,
supraconductivit\'e, etc.

\end{enumerate}
Nous nous contenterons d'\'etudier ici une \'equation plus simple,
celle de Kuramoto et Sivashinsky (KS),
d\'eduite de celle pour la phase $\varphi=\arg \psi$ du champ $\psi$
de CGL3 en prenant une certaine limite \cite{PM1979,Lega2001},
\begin{eqnarray} 
& &
\varphi_t + \nu \varphi_{xxxx} + b \varphi_{xxx} + \mu \varphi_{xx}
 + \varphi \varphi_x = 0,\quad
    \varphi \in \mathcal{C},\,
    (\nu,b,\mu) \in \mathcal{R},\
\nu \not=0.
\label{eqKS}
\end{eqnarray}
Sa r\'eduction en onde propagative
\begin{eqnarray} 
& &
\varphi(x,t)=c+u(\xi),\ \xi=x-ct,
\\
& &
E(u,\xi) \equiv
\nu u''' + b u'' + \mu u' + \frac{u^2}{2} + A = 0,\ 
    (\nu,b,\mu) \in \mathcal{R},\
\nu \not=0,
\label{eqKSODE}
\end{eqnarray}
o\`u $A$ est une constante d'int\'egration,
a \'egalement un comportement chaotique \cite{MannevilleBook},
et elle d\'epend de deux param\`etres sans dimension,
$b^2/(\mu \nu)$ et $\nu A / \mu^3$.
Nous noterons d\'esormais $x$ pour $\xi$.
Le probl\`eme \`a r\'esoudre est le suivant.

\textbf{Probl\`eme}.
Trouver l'expression explicite de la ``solution analytique g\'en\'erale''
(d\'efi\-ni\-tion \textit{supra}) de l'EDO (\ref{eqKSODE}).

La motivation vient de ce que cette solution (pour CGL3 et donc pour KS)
est ``observ\'ee'' sous forme de texture bien d\'efinie,
tant dans des simulations num\'eriques
que dans de v\'eritables exp\'eriences physiques \cite{vHSvS}.

\section{Suppression locale de la contribution chaotique}

Comptons d'abord le nombre (n\'ecessairement inf\'erieur \`a trois)
de constantes arbitraires dans la solution analytique g\'en\'erale.
La recherche d'un comportement local alg\'ebrique
au voisinage d'une singularit\'e mobile $x_0$
(mobile signifie~: qui d\'epend des conditions initiales),
\begin{eqnarray} 
& &
u \sim_{x \to x_0} u_0 \chi^p,\ u_0 \not=0,\ \chi=x-x_0,
\end{eqnarray}
conduit \`a la s\'erie de Laurent 
\begin{eqnarray} 
& &
u^{(0)} = 120 \nu \chi^{-3} - 15b \chi^{-2}
        + \left(\frac{60 \mu}{19} - \frac{15b^2}{76 \nu}\right) \chi^{-1} 
        + O(\chi^0),
\label{eqKSODELaurent}
\end{eqnarray}
d'o\`u sont absentes deux des trois constantes arbitraires.
Celles-ci apparaissent en perturbation \cite{CFP1993}, 
\begin{eqnarray} 
& &
u= u^{(0)} + \varepsilon u^{(1)} + \varepsilon^2 u^{(2)} + \dots,
\end{eqnarray}
o\`u le petit param\`etre $\varepsilon$ ne figure pas dans l'EDO
(\ref{eqKSODE}).
L'\'equation lin\'earis\'ee en $u^{(0)}$
\begin{eqnarray} 
& &
\left(\nu \frac{\D^3}{\D x^3} + b \frac{\D^2}{\D x^2} + \mu \frac{\D}{\D x}
 + u^{(0)}\right) u^{(1)}=0,
\end{eqnarray}
est alors du type de Fuchs au voisinage de $x=x_0$,
avec pour \'equation indicielle
($q=-6$ d\'esigne le degr\'e de singularit\'e du premier membre $E$
de (\ref{eqKSODE}))
\begin{eqnarray}
& & 
\lim_{\chi \to 0} \chi^{-j-q} (\nu \partial_x^3 + u_0 \chi^p) \chi^{j+p}
\\
& & 
= \nu (j-3)(j-4)(j-5) + 120 \nu = \nu (j+1) (j^2 -13 j + 60)
\\
& &
=\nu (j+1) \left(j-\frac{13 + i \sqrt{71}}{2}\right)
           \left(j-\frac{13 - i \sqrt{71}}{2}\right)=0.
\end{eqnarray}
La repr\'esentation locale de la solution g\'en\'erale,
\begin{eqnarray}
u(x_0,\varepsilon c_+,\varepsilon c_-) 
& = & 
 120\nu \chi^{-3} \{ 
 \hbox{Taylor}(\chi) 
\nonumber
\\
& & 
+ \varepsilon [ 
              c_{+} \chi^{(13+i\sqrt{71})/2} \hbox{Taylor}(\chi)
\nonumber
\\
& & 
	    + \ \ c_{-} \chi^{(13-i\sqrt{71})/2} \hbox{Taylor}(\chi) ]
            + {\mathcal O}(\varepsilon^2)\}, 
\nonumber
\end{eqnarray}
o\`u ``Taylor'' d\'esigne des s\'eries convergentes de $\chi$,
d\'epend bien de trois constantes arbitraires 
$(x_0,\varepsilon c_+,\varepsilon c_-)$
(l'indice de Fuchs $-1$ ne repr\'esente qu'une translation de $x_0$).
Le branchement mobile dense provenant des deux indices 
irrationnels caract\'erise \cite{TF} le comportement chaotique,
et l'unique moyen de l'\'eliminer est d'exiger 
$\varepsilon c_+=\varepsilon c_-=0$, \cad\ $\varepsilon=0$,
restreignant ainsi \`a un seul arbitraire 
la partie analytique de la solution.
Le probl\`eme est donc de trouver une expression compacte pour la s\'erie de
Laurent (\ref{eqKSODELaurent}).

\section{Sur la suppression globale de la contribution chaotique}

Par \'elimination de la constante mobile $x_0$ (donc de $x-x_0$) entre
la s\'erie de Laurent (\ref{eqKSODELaurent}) et sa d\'eriv\'ee,
la solution analytique g\'en\'erale satisfait \`a une EDO autonome d'ordre un,
\begin{eqnarray}
& &
F(u',u)=0,
\label{eqsubeq}
\end{eqnarray}
que nous appellerons \textit{sous-\'equation} de (\ref{eqKSODE}),
puisque cette derni\`ere en est une cons\'e\-quen\-ce diff\'eren\-tielle.

Trouver la solution analytique g\'en\'erale \'equivaut alors \`a 
trouver une sous-\'equation autonome d'ordre un.
En effet, si l'on admet que $F$ est polynomiale et que sa solution 
g\'en\'erale est uniforme, alors
\begin{enumerate}
\item
(Hermite \cite[Vol.~II, \S 139]{Valiron})
le genre de la courbe alg\'ebrique $F=0$ est z\'ero ou un,

\item
(Briot et Bouquet \cite[pages 58--59]{PaiLecons})
sa forme n\'ecessaire est 
\begin{eqnarray}
& &
F(u,u') \equiv
 \sum_{k=0}^{m} \sum_{j=0}^{2m-2k} a_{j,k} u^j {u'}^k=0,\ a_{0,m}=1,
\label{eqsubeqODEOrderOnePP}
\end{eqnarray}
o\`u $m$ est un entier positif et les ${a_{j,k}}$ sont des constantes,

\item
(Briot et Bouquet)
la solution est,
pour le genre z\'ero, 
une fraction rationnelle de $e^{a x}$ ou de $x$ 
($a$ d\'esignant une constante);
pour le genre un, une fonction elliptique.

\end{enumerate}

Dans les deux cas, il existe un algorithme, 
implant\'e en Maple par Mark van Hoeij \cite{MapleAlgcurves},
qui donne l'expression explicite de la solution.

Malheureusement, \`a notre connaissance, 
on ne sait pas effectuer l'\'elimination de $x-x_0$
entre la s\'erie de Laurent et sa d\'eriv\'ee qui,
si elle \'etait possible, fournirait (\ref{eqsubeq}).

\section{Rappel sur la classe d'\'equations $P(u^{(n)},u)=0$}

$P$ d\'esignant un polyn\^ome de deux variables \`a coefficients complexes,
la classe d'\'equa\-tions autonomes
\begin{eqnarray}
& &
P(u^{(n)},u)=0,
\label{eqODEDeponent}
\end{eqnarray}
ind\'ependantes des d\'eriv\'ees d'ordre $2$ \`a $n-1$ inclus,
a \'et\'e classiquement beaucoup \'etudi\'ee.
Soit $W$ la classe de fonctions d'une variable complexe $x$
\begin{eqnarray}
& &
W:=
\lbrace\hbox{rationnel}(x),\hbox{rationnel}(e^{a x}),\hbox{elliptique}(x)
\rbrace,
\label{eqClassW}
\end{eqnarray}
o\`u $a$ est une constante complexe.
C'est en fait la classe des fonctions elliptiques et de leurs 
d\'eg\'en\'erescences, car
\begin{eqnarray}
& &
\lim_{a \to 0}\hbox{rationnel}(e^{a x})=
\lim_{a \to 0}\hbox{rationnel}((e^{a x}-1)/a)=
              \hbox{rationnel}(x).
\end{eqnarray}
Les r\'esultats connus sont les suivants.
\begin{enumerate}
\item
(Briot et Bouquet)
Pour $n=1$, si la solution est uniforme, alors elle est dans $W$.

\item
(Picard)
Pour $n=2$, toute solution m\'eromorphe est dans $W$.

\item
(Eremenko \cite{Eremenko1986})
Si $n$ est impair,
si le genre de la courbe alg\'ebrique $P=0$ est nul,
et s'il existe une solution particuli\`ere m\'eromorphe qui poss\`ede au
moins un p\^ole,
alors cette solution est dans $W$.

\end{enumerate}

Ce troisi\`eme cas (qui se d\'emontre par la th\'eorie de Nevanlinna)
s'applique \`a (\ref{eqKSODE}) pour $b=\mu=0$,
la solution est alors connue \cite{FournierSpiegelThual}
\begin{eqnarray}
& &
b=0,\ \mu=0\ :\ 
u = -60\nu \wp'(x-x_0,0,g_3),\ g_2=0,\ g_3=\frac{A}{1080 \nu^2},
\end{eqnarray}
$\wp$ repr\'esentant la fonction elliptique de Weierstrass,
et il en existe m\^eme une extrapolation de codimension un \cite{Kud1990}
\begin{eqnarray}
& &
{\hskip -10.0 truemm}
b^2 - 16 \mu \nu=0\ :\ 
u=-60\nu \wp' - 15b\wp - \frac{b \mu}{4 \nu},\
g_2=\frac{\mu^2}{12 \nu^2},\
g_3=\frac{13\mu^3+\nu A}{1080 \nu^3}.
\label{eqKSConstraint16}
\label{eqKSElliptic}
\end{eqnarray}
Notre probl\`eme revient \`a extrapoler cette solution
pour une valeur arbitraire de $b^2/(\mu \nu)$.

\section{Solutions de codimension non nulle}

Elles appartiennent toutes (du moins celles qui sont connues)
\`a la classe $W$, Eq.~(\ref{eqClassW}).
L'unique solution connue de codimension trois est rationnelle,
\begin{eqnarray}
& &
b=0,\ \mu=0,\ A=0\:\ u = 120\nu (x-x_0)^{-3}.
\end{eqnarray}
On conna\^{\i}t six solutions de codimension deux
\cite{KuramotoTsuzuki,KudryashovKSFourb},
toutes rationnelles en $e^{k x}$, 
\begin{eqnarray}
u & = &
 \frac{5}{2} b k^2 - \frac{13 b^3}{32 \times 19 \nu^2}
  + \frac{7 \mu b}{4 \times 19 \nu}
 +\left(\frac{60}{19} \mu - 30 \nu k^2 - \frac{15 b^2}{4 \times 19 \nu}\right)
       \frac{k}{2} \tanh \frac{k}{2} (x-x_0)
\nonumber
\\
& &
 - 15 b
 \left(\frac{k}{2} \tanh \frac{k}{2} (x-x_0)\right)^2
 + 120 \nu
 \left(\frac{k}{2} \tanh \frac{k}{2} (x-x_0)\right)^3,\
\label{eqKSTrigo}
\end{eqnarray}
les valeurs permises figurant dans la Table \ref{Table1}.

\begin{table}[h] 
\caption[garbage]{
Les six solutions connues de codimension deux de (\ref{eqKSODE}).
Elles ont toutes la forme (\ref{eqKSTrigo}).
}
\vspace{0.2truecm}
\begin{center}
\begin{tabular}{| c | c | c |}
\hline 
\hline 
$b^2/(\mu\nu)$ & $\nu A/\mu^3$ & $\nu k^2/\mu$
\\ \hline \hline 
$0$ & $-4950/19^3,\ 450/19^3$ & $11/19,\ -1/19$
\\ \hline 
$144/47$ & $-1800/47^3$ & $1/47$
\\ \hline 
$256/73$ & $-4050/73^3$ & $1/73$
\\ \hline 
$16$ & $-18,\ -8$ & $1,\ -1$
\\ \hline 
\end{tabular}
\end{center}
\label{Table1}
\end{table}

Enfin, l'unique solution connue de codimension un est elliptique, 
cf.~(\ref{eqKSElliptic}),
et c'est une extrapolation de la derni\`ere ligne de la Table \ref{Table1}.

Une belle propri\'et\'e commune \`a toutes ces solutions est d'admettre la
repr\'esenta\-tion
\begin{eqnarray}
u & = &
\mathcal{D} \Log \sigma,\
\mathcal{D}=60 \nu \frac{\D }{\D x^3} + 15 b \frac{\D }{\D x^2}
           + \frac{15(16 \mu \nu - b^2)}{76 \nu} \frac{\D }{\D x},
\label{eqKSD}
\end{eqnarray}
o\`u $\sigma$ est une fonction enti\`ere dont l'EDO est facile \`a construire.

Avant de rechercher plus avant la solution inconnue de codimension z\'ero,
il convient de se poser la question si elle est ou non dans la classe $W$,
c.\`a.d.~si elle est elliptique.

On pourrait craindre en effet que cette classe $W$ soit insuffisante,
au vu d'une EDO qui ne diff\`ere de (\ref{eqKSODE}) que par le terme $u u'$
\cite{Samsonov1994,Porubov1996},
\begin{eqnarray}
& &
{\hskip -10.0 truemm}
E \equiv 
 -u''' + 9 k u'' +(a_1  u     -26 k^2) u' -2 a_1 k  u     ^2 + 24 k^2  u,
\end{eqnarray}
et dont la solution sort de la classe $W$,
\begin{eqnarray}
& &
{\hskip -10.0 truemm}
u= \frac{12}{a_1} e^{2 k x} \wp\left(\frac{e^{k x}-1}{k}-X_0,g_2,g_3\right),\
(X_0,g_2,g_3) \hbox{ arbitraires}.
\label{eqSamsonovSol}
\end{eqnarray}
Une solution dans cette nouvelle classe est exclue pour (\ref{eqKSODE}),
car l'EDO d'ordre un r\'esultant de l'\'elimination de $X_0$ entre
(\ref{eqSamsonovSol}) et sa d\'eriv\'ee,
\begin{eqnarray}
& &
(u'-2 k u)^2 - 4 u^3 + e^{4 k x} g_2 u + e^{6 k x} g_3=0,
\end{eqnarray}
n'est autonome que pour la d\'eg\'en\'erescence rationnelle $g_2=g_3=0$.

\section{Vers une approche num\'erique de la solution elliptique}

\`A partir de l'unique \'el\'ement d'information que constitue la s\'erie de
Laurent (\ref{eqKSODELaurent}),
nous allons d\'efinir plusieurs approches num\'eriques
pour tester la validit\'e de l'hypoth\`ese elliptique
(appartenance \`a la classe $W$).
L'outil naturel est celui des approximants de Pad\'e (\AP) \cite{PadeIntro}.

\'Etant donn\'e les $N$ premiers termes d'une s\'erie de Taylor en $x=0$,
\begin{eqnarray}
& &
S_N = \sum\limits_{j=0}^N c_j x^j,
\end{eqnarray}
l'approximant de Pad\'e $[L,M]$ de la s\'erie est d\'efini comme
l'unique fraction rationnelle
\begin{eqnarray}
& &
[L,M] = \frac{\sum_{l=0}^L a_l x^l}{\sum_{m=0}^M b_m x^m},\ b_0=1,
\end{eqnarray}
satisfaisant \`a la condition
\begin{eqnarray}
& &
S_N - [L,M] = {\mathcal O}(x^{N+1}),\qquad L+M=N.
\end{eqnarray}
L'extension aux s\'eries de Laurent ne pr\'esente pas de difficult\'e.
En particulier, les \AP\ sont exacts sur les fractions rationnelles
d\`es que $L$ et $M$ sont suffisamment grands.

Le calcul pr\'ealable d'un grand nombre (environ $200$) de termes de la 
s\'erie de Laurent se fait par un algorithme lin\'eaire,
la seule difficult\'e, ais\'ee \`a surmonter, \'etant de ne jamais engendrer
de termes qui s'av\'ereront inutiles.
Il importe \'egalement de calculer les $c_j$, puis les $(a_l,b_m)$,
sur $\mathcal Q$, jamais sur $\mathcal R$,
\`a cause d'importantes compensations.
C'est pour cela que nous avons d\^u \'ecarter le programme \textit{pade} de
\textit{Numerical recipes} \cite{NumericalRecipes},
qui engendre des doublets de Froissart \cite{GP1997,GP1999} ind\'esirables
(couples d'un z\'ero et d'un p\^ole tr\`es proches,
ces couples se r\'epartissant sur un cercle sans d\'efinir de coupure
comme ce serait le cas s'il ne s'agissait pas d'erreurs num\'eriques).
La fonction \textit{pade} du langage formel Maple \cite{Maple} 
est en revanche tout \`a fait adapt\'ee.

\textit{Remarque}.
Il est possible d'\'eviter l'\'etape de calcul des \AP\
et d'obtenir directement et rapidement \cite{TY} les valeurs num\'eriques
des z\'eros et des p\^oles du Pad\'e $[L,M]$.

\subsection{Premier test d'ellipticit\'e}

La structure des p\^oles et des z\'eros des \AP\ de (\ref{eqKSODELaurent})
caract\'erise la classe de fonctions \`a laquelle appartient sa somme 
inconnue.

Une somme elliptique est caract\'eris\'ee 
par une double infinit\'e de p\^oles et de z\'eros
situ\'es aux n{\oe}uds d'un r\'eseau doublement p\'eriodique,
une somme rationnelle en $e^{a x}$, avec $a$ complexe,
par une simple infinit\'e de p\^oles et de z\'eros,
une somme rationnelle en $x$
par un nombre fini de p\^oles et de z\'eros, etc.
C'est pour tester les structures doublement p\'eriodiques que l'on a besoin
d'un grand nombre de termes dans la s\'erie de Laurent.

\begin{figure}[ht]
\centerline{
 \epsfxsize=2.4in 
 \epsfbox{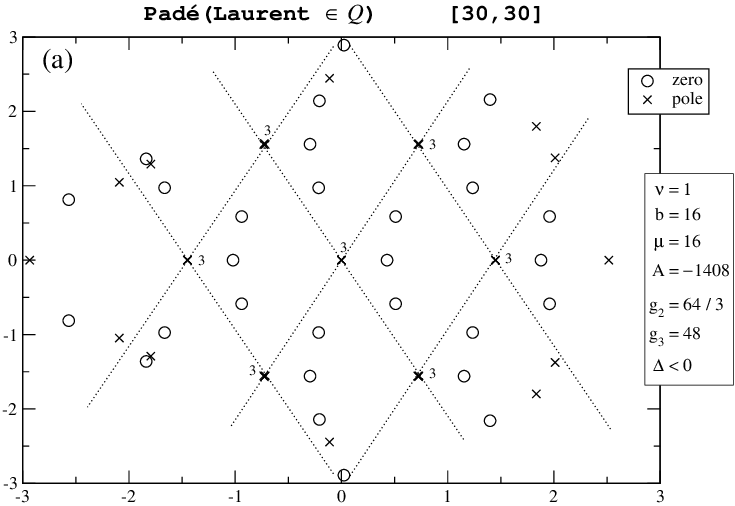} 
 \qquad
 \epsfxsize=2.4in
 \epsfbox{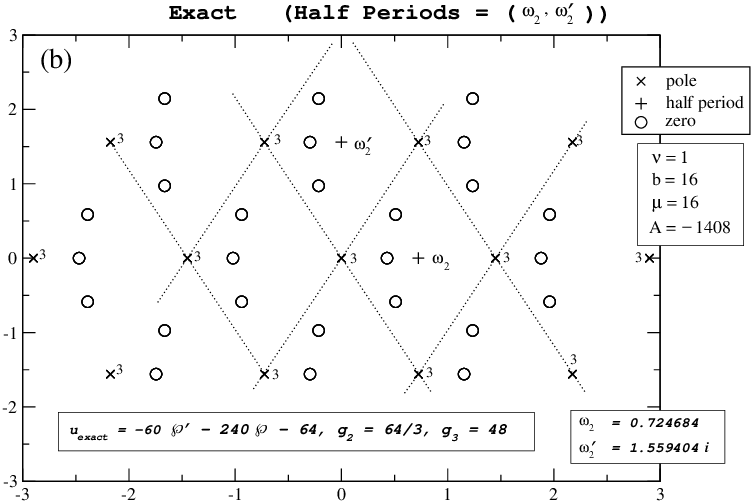}
 }
 \vspace{0.5cm}
\centerline{ \epsfxsize=2.4in \epsfbox{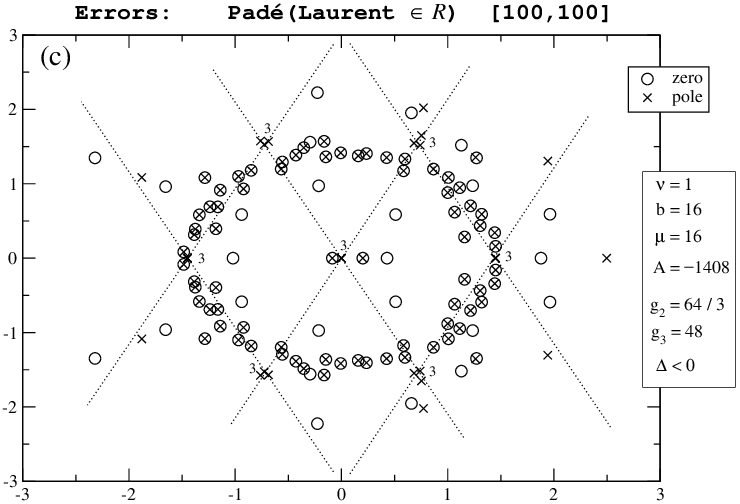}  \qquad
             \epsfxsize=2.4in \epsfbox{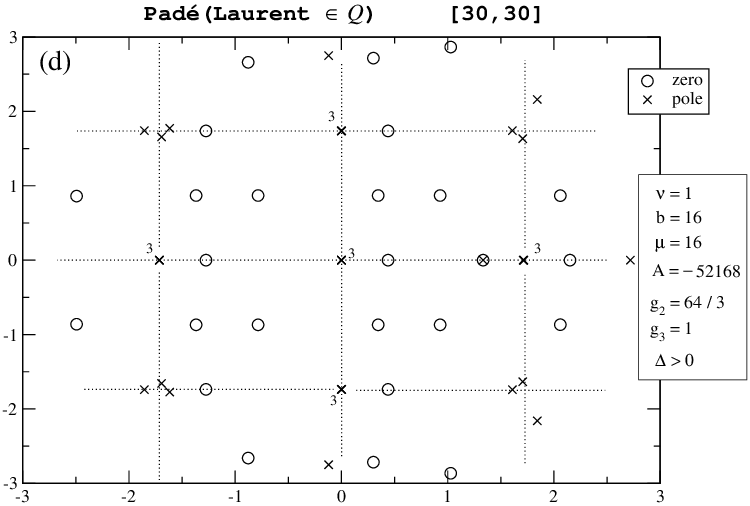}}
\caption{
Plan complexe $x$ du seul cas elliptique connu 
($b^2=16\mu\nu$, $A=\mbox{arb}$).
La structure des z\'eros (symbole $0$) et des p\^oles (symbole $X$) 
doit \^etre doublement p\'eriodique,
avec trois z\'eros simples et un p\^ole triple par 
parall\'elogramme p\'eriode.
Figure (d)~: $(\nu,b,\mu,A)=(1,16,16,-52168)$,
calcul de l'\AP\ $[30,30]$ sur $\mathcal Q$.
Les trois autres figures sont trac\'ees pour $(\nu,b,\mu,A)=(1,16,16,-1408)$,
donc $(g_2,g_3)=(64/3,48)$.
Figure (b)~: emplacements exacts des z\'eros et des p\^oles,
calcul\'es \`a partir de la solution exacte (\ref{eqKSElliptic}).
Figure (c)~: z\'eros et p\^oles de l'\AP\ $[100,100]$ 
calcul\'e sur $\mathcal R$,
montrant l'accumulation des doublets de Froissart sur le cercle unit\'e
(\'echelles diff\'erentes sur les deux axes).
Figure (a)~: z\'eros et p\^oles de l'\AP\ $[30,30]$ 
calcul\'e sur $\mathcal Q$.
Dans tous les cas,
on discerne nettement les trois z\'eros simples et le p\^ole triple
de chaque p\'eriode.
}
\label{FigKSPlanxEllipticKnown}
\end{figure}

\begin{figure}[ht]
\centerline{ \epsfxsize=2.4in \epsfbox{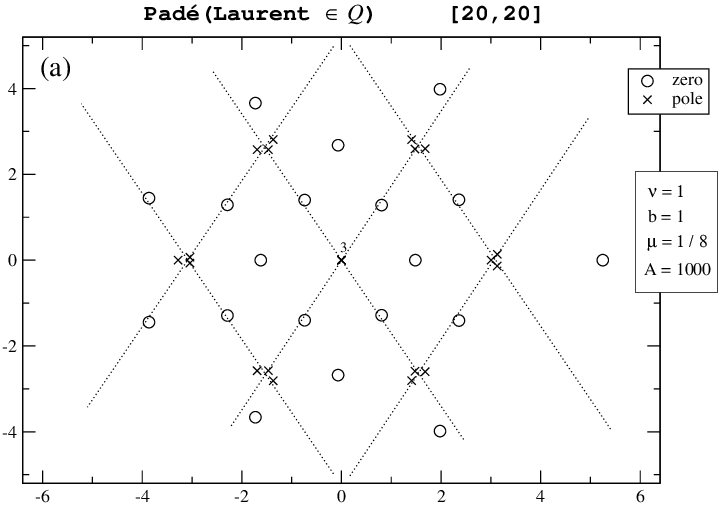}  \qquad
             \epsfxsize=2.4in \epsfbox{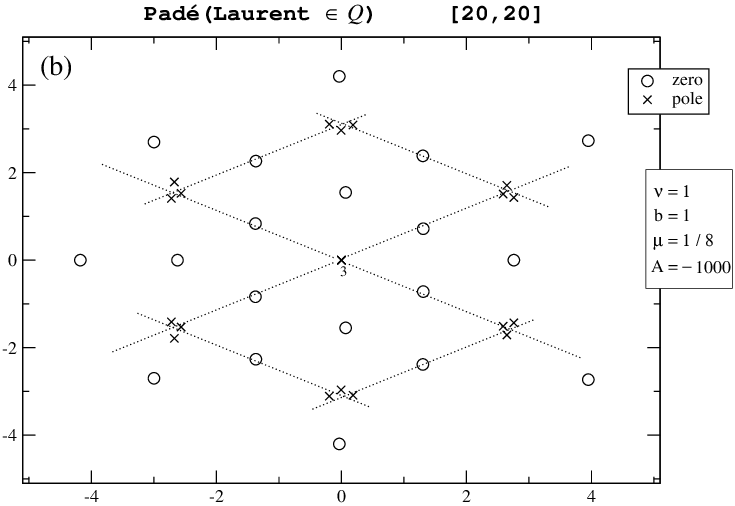}} \vspace{0.5cm}
\centerline{ \epsfxsize=2.4in \epsfbox{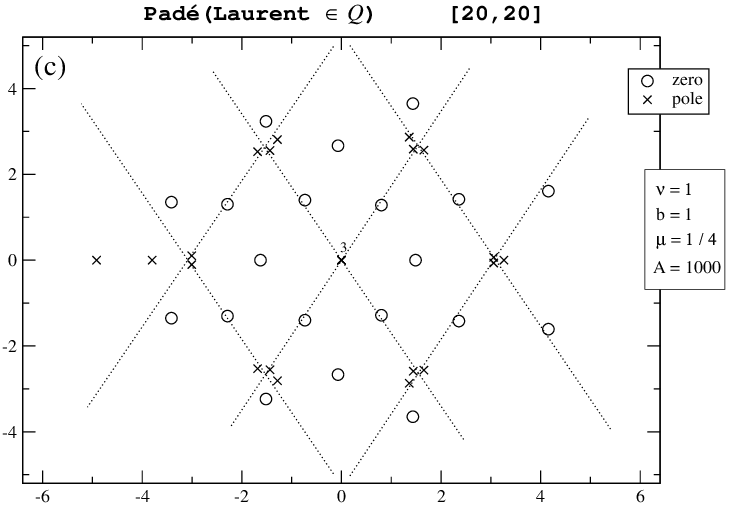}  \qquad
             \epsfxsize=2.4in \epsfbox{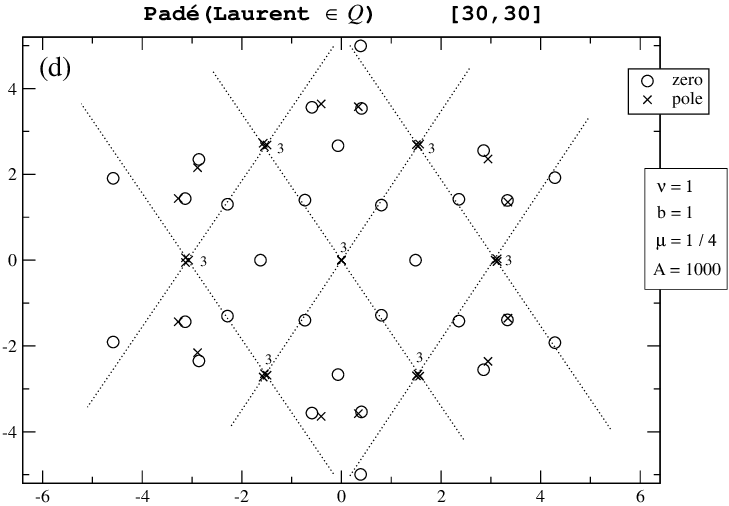}}
\caption{
Plan complexe $x$ pour les valeurs $b^2/(\mu \nu)=4$ et $8$
(cas g\'en\'eriques inconnus)
et des \AP\ calcul\'es sur $\mathcal Q$.
Fig.~(a), $(\nu,b,\mu)=(1,1,1/8),A=1000$.
Fig.~(b), $(\nu,b,\mu)=(1,1,1/8),A=-1000$.
Figs.~(c) et (d), $(\nu,b,\mu,A)=(1,1,1/4,1000)$,
pour $[L,M]=[20,20]$ (Fig.~(c)) et $[30,30]$ (Fig.~(d)).
La structure est, avec une bonne pr\'ecision, doublement p\'eriodique
\`a trois z\'eros simples et un p\^ole triple par p\'eriode.
}
\label{FigKSPlanxUnknown}
\end{figure}

Pour tous les cas de solution connue 
((\ref{eqKSTrigo}), (\ref{eqKSElliptic})),
en choisissant pour les constantes fixes $(\nu,b,\mu,A)$ des valeurs 
rationnelles,
le r\'esultat est conforme aux esp\'erances, 
voir Fig.~\ref{FigKSPlanxEllipticKnown}.
Pour plusieurs cas g\'en\'eriques de valeurs de $b^2/(\mu\nu)$
(par exemple $b^2/(\mu\nu) = 4, 8$, 
Fig.~\ref{FigKSPlanxUnknown}),
nous trouvons une solide indication num\'erique que
les textures observ\'ees sont toutes doublement p\'eriodiques,
ce qui est la signature des fonctions elliptiques.
Dans tous les cas, connus et inconnus,
il est ais\'e de compter le nombre (commun) de z\'eros et de p\^oles par 
p\'eriode, et les calculs indiquent que ce nombre serait \'egal \`a trois.

\textit{Remarque}.
La connaissance du nombre et des positions des z\'eros et des p\^oles 
d'une p\'eriode permet la d\'etermination num\'erique des 
invariants $g_2$ et $g_3$.

\subsection{Deuxi\`eme test d'ellipticit\'e}

Si la solution $u$ est elliptique, alors, d'apr\`es un r\'esultat classique,
il existe deux fractions rationnelles $R_1, R_2$ telles que
$u = R_1(\wp) + R_2(\wp)\,\wp'$, plus pr\'ecis\'ement
\begin{equation} 
u(x,g_2,g_3) = \underbrace{\frac{\mbox{Poly}_{N_1}(\wp)}
                {\mbox{Poly}_{D}(\wp)}}_{\mbox{\small partie paire}}
                + \frac{\mbox{Poly}_{N_2}(\wp)}{\mbox{Poly}_{D}(\wp)}
                \sqrt{4\wp^3 - g_2\wp - g_3},
\label{eqRationalwpwpprime}
\end{equation}
o\`u $\mbox{Poly}_{D}$ d\'esigne un polyn\^ome de degr\'e $D$.

En vue d'une recherche directe de la solution sous la forme 
(\ref{eqRationalwpwpprime}),
il est en principe possible de calculer num\'eriquement les degr\'es 
$N_1,N_2,D$ des fractions rationnelles de $\wp$,
en rempla\c cant le plan complexe de $x$ par celui de $\wp(x)$,
de la mani\`ere suivante.

Tout d'abord, la n\'ecessit\'e d'une seule variable complexe dans les \AP\
oblige \`a \'eliminer $\wp'$, donc \`a introduire le branchement alg\'ebrique
ind\'esirable $\sqrt{4\wp^3 - g_2\wp - g_3}$.
Un moyen simple de le supprimer est alors, 
puisque $\wp$ est une fonction paire,
de se restreindre \`a la partie paire de la s\'erie de Laurent 
$u(\chi=x-x_0)$,
\begin{eqnarray}
& &
u^{\rm paire} = 
 - 15b \chi^{-2}
 + \frac{b (56 \mu \nu - 13 b^2)}{19 \times 32\nu^2}
 - \frac{b \left(10\mu\nu-3b^2\right)^2}{19^2 \times 64 \nu^4} \chi^2
 + {\mathcal O}(\chi^4),
\label{eqKSODELaurentEven}
\end{eqnarray}
d'inverser \cite[vol.~II, p.~527]{Knuth} la s\'erie de Laurent $\wp$ de $x$,
\begin{eqnarray}
& &
\wp (x,g_2,g_3) = x^{-2} + \frac{g_2}{20} x^2 + \frac{g_3}{28} x^4
 + {\mathcal O}(x^6)
\end{eqnarray}
en la s\'erie de Laurent $x^2$ de $\wp$,
\begin{eqnarray}
x^2 & = & \wp^{-1} \left[ 1 + \frac{g_2}{20}\, \wp^{-2}
        + \frac{g_3}{28}\, \wp^{-3}
        + \frac{7 g_2^2}{1200}\, \wp^{-4}
        + \frac{29 g_2 g_3}{3080}\, \wp^{-5} \right.
\nonumber\\  
        & & \hspace{0.8cm}
        + \left(\frac{11 g_2^3}{12480} + \frac{5 g_3^2}{1274}\right) \wp^{-6}
        + \frac{167 g_2^2 g_3}{73920}\, \wp^{-7} 
\nonumber\\
        & & \hspace{0.8cm} \left.
        + \left(\frac{77 g_2^4}{509184} + \frac{669 g_2 g_3^2}{340340}\right)
        \wp^{-8} + {\mathcal O}(\wp^{-9}) \right], 
\label{eqLaurentx2wp}
\end{eqnarray}
puis de composer les deux s\'eries 
(\ref{eqKSODELaurentEven}) et (\ref{eqLaurentx2wp}) pour obtenir
\begin{eqnarray}
u^{\rm paire}
& = &
-15 b \wp
 + \frac{b (56 \mu \nu - 13 b^2)}{19 \times 32\nu^2}
 + \left( \frac{3 b g_2}{4}
 - \frac{b \left(10\mu\nu-3b^2\right)^2}{19^2 \times 64 \nu^4} 
\right) \wp^{-1}
\nonumber
\\
& &
 + {\mathcal O}(\wp^{-2}).
\label{eqLaurentuevenwp}
\end{eqnarray}

La question est alors de tester si cette s\'erie (\ref{eqLaurentuevenwp}),
dont les coefficients d\'ependent de $(\nu,b,\mu,A,g_2,g_3)$,
a bien pour somme une fraction rationnelle
\begin{equation} 
u^{\rm paire}(x,g_2,g_3)
 = \frac{\mbox{Poly}_{N_1}(\wp)}{\mbox{Poly}_{D}(\wp)}.
\label{equeven}
\end{equation}

\begin{figure}[ht]
\centerline{ \epsfxsize=2.4in \epsfbox{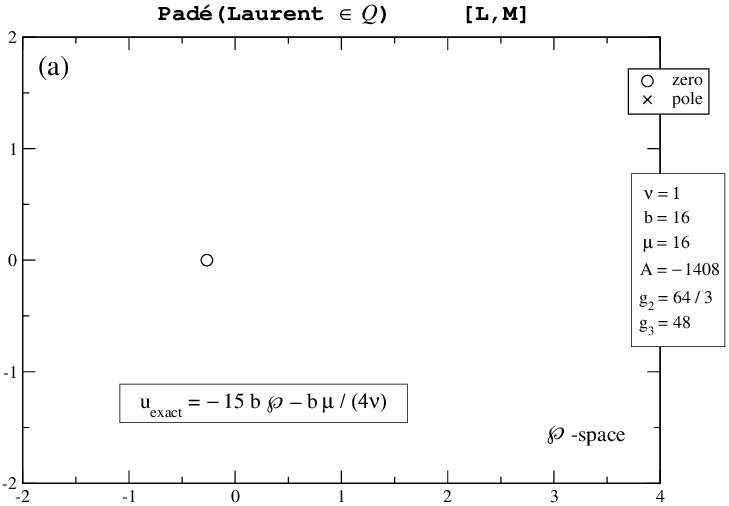}  \qquad
             \epsfxsize=2.4in \epsfbox{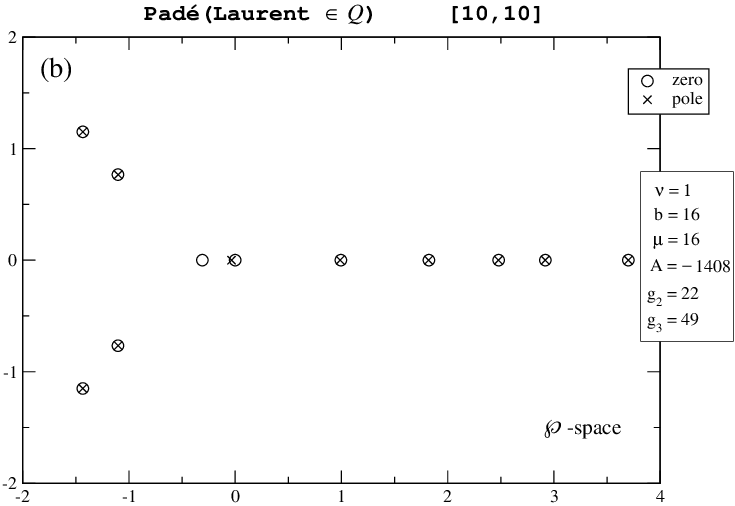}}
\caption{
Plan complexe $\wp$ de la partie paire de la solution $u$
pour $b^2/(\mu \nu)=16, A=-1408$,
ce qui d\'efinit par (\ref{eqKSElliptic}) les valeurs $g_2=64/3,g_3=48$.
Dans la figure (a),
$(\nu,b,\mu,A,g_2,g_3)=(1,16,16,-1408,64/3,48)$,
tout \AP\ $[L,M]$ est exact car calcul\'e sur $\mathcal Q$,
la figure contient donc l'unique z\'ero $\wp=- \mu /(60 \nu)$
et elle n'a aucun p\^ole.
Dans la figure (b),
o\`u les valeurs de $(g_2,g_3)$ sont choisies rationnelles et l\'eg\`erement
diff\'erentes des valeurs exactes,
$(\nu,b,\mu,A,g_2,g_3)=(1,16,16,-1408,22,49)$,
l'\AP\ $[10,10]$, calcul\'e sur $\mathcal Q$,
fait constater $10$ quasi-compensations d'un p\^ole par un z\'ero.
}
\label{FigKSPlanwpEllipticKnown}
\end{figure}

La difficult\'e provient de la n\'ecessit\'e de donner,
dans la s\'erie (\ref{eqLaurentuevenwp}),
des valeurs num\'eriques, de pr\'ef\'erence rationnelles,
\`a $(g_2,g_3)$ avant de pouvoir calculer les 
z\'eros et les p\^oles des \AP.
Puisque le parall\'elogramme p\'eriode du plan $x$
est d\'ej\`a d\'etermin\'e,
il permet de calculer des valeurs approch\'ees (complexes) de $(g_2,g_3)$.
Il importe donc d'\'etudier la sensibilit\'e du r\'esultat 
(valeurs de $N_1,D$) aux valeurs de $(g_2,g_3)$,
et cette sensibilit\'e ne peut \^etre test\'ee que sur
l'unique solution elliptique connue (\ref{eqKSElliptic}).
La Figure \ref{FigKSPlanwpEllipticKnown} 
montre la structure des \AP\ de (\ref{eqLaurentuevenwp}) dans le plan
complexe $\wp$ pour $b^2=16 \mu \nu$
et pour un choix de $(g_2,g_3)$ voisin de celui de ce cas.

\subsection{D\'etermination des degr\'es de la sous-\'equation}

Suivant une suggestion de Belabas faite lors de la conf\'erence,
les degr\'es de la courbe alg\'ebrique (\ref{eqsubeqODEOrderOnePP})
pourraient se calculer par une technique de Pad\'e, comme suit.
\`A partir de la s\'erie de Laurent (\ref{eqKSODELaurent}),
on construit d'abord une autre s\'erie de Laurent \`a p\^ole simple,
par exemple $u'/u=\hbox{Laurent}(x-x_0)$,
on l'inverse en $x-x_0=\hbox{Laurent}(u'/u)$,
et on la substitue dans la s\'erie $u=\hbox{Laurent}(x-x_0)$.
La s\'erie r\'esultante $u=\hbox{Laurent}(u'/u)$,
qui r\'ealise sous forme locale l'\'elimination de $x-x_0$ entre $u$ et $u'$,
est alors analys\'ee par Pad\'e.
Du branchement alg\'ebrique ainsi trouv\'e,
il devrait \^etre possible de d\'eduire le degr\'e en $U$ de la
courbe alg\'ebrique $G(U,V)=0$, avec $U=u,V=u'/u$,
donc les degr\'es de (\ref{eqsubeqODEOrderOnePP}).

\section{Approches analytiques}

L'\'etude num\'erique pr\'ec\'edente semble conclure \`a la nature
elliptique de la solution cherch\'ee.
Nous pr\'esentons maintenant bri\`evement trois m\'ethodes analytiques
susceptibles de fournir les coefficients de la sous-\'equation 
(\ref{eqsubeq}).
Si de plus celle-ci est suppos\'ee alg\'ebrique et \`a solution
g\'en\'erale uniforme,
alors elle a la forme n\'ecessaire (\ref{eqsubeqODEOrderOnePP}).

\subsection{\'Equation de Monge-Amp\`ere \'equivalente}

La fonction inconnue de deux variables $F(u,u')$ qui d\'efinit la
sous-\'equation d'ordre un (\ref{eqsubeq})
satisfait \`a une \'equation aux d\'eriv\'ees partielles 
tr\`es simple,
obtenue par \'elimination de $u''$ et $u'''$ entre (\ref{eqKSODE})
et les deux \'equations
\begin{eqnarray}
& &
\frac{\D}{\D x} F(u(x),u'(x))=0,
\\
& &
\frac{\D^2}{\D x^2} F(u(x),u'(x))=0.
\end{eqnarray}
Dans la notation classique
\begin{eqnarray}
& &
X=u,\
Y=u',\
P=\frac{\partial   F}{\partial X},\
Q=\frac{\partial   F}{\partial Y},\
\nonumber
\\
& &
R=\frac{\partial^2 F}{\partial X^2},\
S=\frac{\partial^2 F}{\partial X \partial Y},\
T=\frac{\partial^2 F}{\partial Y^2},\
\end{eqnarray}
c'est une \'equation de Monge-Amp\`ere qui s'\'ecrit
\begin{eqnarray}
& &
{\hskip -13.0 truemm}
 \nu \frac{Y^2(-Q^2 R + 2 P Q S - P^2 T) + Y P^2 Q}{Q^3}
- b  Y \frac{P}{Q} + \mu Y + \frac{X^2}{2} + A =0.
\label{eqMomgeAmpere}
\end{eqnarray}
Notre probl\`eme \'equivaut alors \`a en trouver une solution particuli\`ere
\`a z\'ero param\`etre qui soit polynomiale.
Ce nouveau probl\`eme pourrait n'\^etre pas plus simple que l'original,
car la m\'ethode des caract\'eristiques est inapplicable puisque
l'\'equation (\ref{eqMomgeAmpere}) h\'erite de la nature chaotique de KS.
En particulier nous n'avons pas r\'eussi \`a obtenir les degr\'es
de cette solution polynomiale, si elle existe.

\subsection{Construction de la sous-\'equation par l'algorithme LLL}

Il existe une m\'ethode \cite{lll1982,js1998}, 
h\'erit\'ee de la cryptographie,
qui fournit \textit{en nombres entiers} les coefficients de 
(\ref{eqsubeqODEOrderOnePP}) pour $m$ donn\'e
(et $a_{0,m}$ non normalis\'e \`a un).
Connue sous le nom d'\textit{algorithme de r\'eduction du r\'eseau}
(lattice reduction algorithm) 
ou d'\textit{algorithme LLL},
elle re\c coit en entr\'ee
l'entier $m$,
la suite des inconnues $a_{j,k}$ de (\ref{eqsubeqODEOrderOnePP}),
et $L$ valeurs num\'eriques $(u_i,u_i'),i=1,\cdots,L$
(donc dans $\mathcal R$ et non pas dans $\mathcal Q$)
de la somme des s\'eries de Laurent $u$ et $u'$
aux points $x_i-x_0, i=1,\cdots,L$.
En sortie,
si $L$ est suffisamment grand devant le nombre $(m+1)^2$ d'inconnues
$a_{j,k}$,
elle exhibe en un temps polynomial une suite d'entiers $a_{j,k}$,
par r\'esolution du syst\`eme lin\'eaire homog\`ene surd\'etermin\'e
\begin{eqnarray}
& &
\forall l=1,\dots,L\ :\
 \sum_{k=0}^{m} \sum_{j=0}^{2m-2k} a_{j,k} u_l^j {u_l'}^k=0.
\end{eqnarray}
Si les entiers ainsi trouv\'es sont tous nuls, il faut augmenter $m$ et
recommencer.
Nous n'avons pas encore mis cette m\'ethode en {\oe}uvre car celle
de la section suivante semble beaucoup plus prometteuse.

\subsection{Construction de la sous-\'equation par la s\'erie de Laurent}

Deux d'entre nous ont r\'ecemment propos\'e \cite{MC2003} 
une m\'ethode analytique
permettant de construire la sous-\'equation (\ref{eqsubeqODEOrderOnePP}) 
\`a partir de la s\'erie de Laurent (\ref{eqKSODELaurent}).
L'algorithme est le suivant.

\begin{enumerate}

\item
Choisir un entier positif $m$ et d\'efinir l'EDO de Briot and Bouquet
(\ref{eqsubeqODEOrderOnePP}).
Elle contient $(m+1)^2-1$ constantes inconnues $a_{j,k}$.

\item
Calculer $\jmax$ termes de la s\'erie de Laurent,
o\`u     $\jmax$ est l\'eg\`erement sup\'erieur \`a $(m+1)^2-1$,
\begin{eqnarray}
& &
u=\chi^p
 \left(\sum_{j=0}^{\jmax} u_j \chi^j + {\mathcal O}(\chi^{j+1})\right),
\label{eqLaurent}
\end{eqnarray}
avec dans le cas de KS $p=-3$.

\item
Exiger que la s\'erie de Laurent satisfasse \`a l'EDO de Briot and Bouquet,
c.\`a.d.~demander l'annulation identique de la s\'erie de Laurent 
de $F(u,u')$ \`a l'ordre $\jmax$
\begin{eqnarray}
& &
F \equiv \chi^{D}
 \left(\sum_{j=0}^{\jmax} F_j \chi^j + {\mathcal O}(\chi^{\jmax+1})\right),\
D=m(p-1),\
\forall j\ : \ F_j=0.
\end{eqnarray}
S'il n'existe pas de solution pour $a_{j,k}$,
augmenter $m$ et revenir \`a l'\'etape 1.

\item
Pour toute solution, 
int\'egrer l'EDO autonome d'ordre un (\ref{eqsubeqODEOrderOnePP}).

\end{enumerate}

Le c{\oe}ur de la m\'ethode est la troisi\`eme \'etape,
o\`u le syst\`eme de $\jmax+1$ \'equations $F_j=0$ 
aux $(m+1)^2-1$ inconnues $a_{j,k}$
est \textit{lin\'eaire} and \textit{surd\'etermin\'e},
donc tr\`es facile \`a r\'esoudre.

Quant \`a la quatri\`eme \'etape,
elle est r\'esolue par le progiciel \textit{algcurves} \cite{MapleAlgcurves},
qui met en {\oe}uvre des algorithmes de Poincar\'e.

Dans le cas de l'\'equation (\ref{eqKSODE}) \'etudi\'ee ici,
du fait de la valeur $p=-3$,
la valeur minimale de $m$ est $m=3$,
ce qui correspond \`a $15$ inconnues $a_{j,k}$
et m\^eme \`a $10$ inconnues seulement,
\`a cause de la r\`egle de s\'election suppl\'ementaire
$m(p-1) \le j p + k (p-1)$, 
qui provient du degr\'e de singularit\'e $p=-3$ de $u$.
Ces dix coefficients sont d\'etermin\'es par le syst\`eme de Cramer
d\'efini par les dix \'equations 
$F_j=0,j=0:6,8,9,12$.
Le syst\`eme r\'esiduel,
qui est non-lin\'eaire et surd\'etermin\'e aux inconnues fixes
$(\nu,b,\mu,A)$,
a pour PGCD $b^2-16 \mu \nu$ 
(cf.~(\ref{eqKSConstraint16})),
ce qui d\'efinit la solution de codimension un,
\begin{eqnarray}
& &
\frac{b^2}{\mu \nu}=16,\
\left(u' + \frac{b}{2 \nu} u_s\right)^2
\left(u' - \frac{b}{4 \nu} u_s\right)
+\frac{9}{40 \nu}
\left(u_s^2 + \frac{60 \mu^3}{\nu} + \frac{10 A}{3}\right)^2=0,\
\nonumber
\\
& &
u_s=u+\frac{3 b \mu}{2 \nu}\cdot
\label{eqKSsubeqgenus1}
\end{eqnarray}
Apr\`es division par ce PGCD,
le syst\`eme restant aux inconnues $(\nu,b,\mu,A)$
admet exactement quatre solutions
(il est suffisant d'arr\^eter la s\'erie \`a $j=16$ pour obtenir ce 
r\'esultat),
\begin{eqnarray}
& &
{\hskip -10.0 truemm}
b=0,\
\nonumber
\\
& &
{\hskip -10.0 truemm}
\left(u' + \frac{180 \mu^2}{19^2 \nu}\right)^2 
\left(u' - \frac{360 \mu^2}{19^2 \nu}\right)
+\frac{9}{40 \nu}
\left(u^2 + \frac{30 \mu}{19} u' - \frac{30^2 \mu^3}{19^2 \nu}\right)^2=0,\
\label{eqKSsubeqgenus0first}
\\
& &
{\hskip -10.0 truemm}
b=0,\
{u'}^3
+\frac{9}{40 \nu}
\left(u^2 + \frac{30 \mu}{19} u' + \frac{30^2 \mu^3}{19^3 \nu}\right)^2=0,\
\\
& &
{\hskip -10.0 truemm}
\frac{b^2}{\mu \nu}=\frac{144}{47},\
u_s=u-\frac{5 b \mu}{47 \nu},\
\left(u' + \frac{b}{4 \nu} u_s\right)^3+\frac{9}{40 \nu} u_s^4=0,\
\\
& &
{\hskip -10.0 truemm}
\frac{b^2}{\mu \nu}=\frac{256}{73},\
u_s=u-\frac{45 b \mu}{584 \nu},\
\nonumber
\\
& &
{\hskip -10.0 truemm}
\left(u' + \frac{b}{8 \nu} u_s\right)^2
\left(u' + \frac{b}{2 \nu} u_s\right)
+\frac{9}{40 \nu}
\left(u_s^2+\frac{5 b^3}{1024 \nu^2}u_s + \frac{5 b^2}{128 \nu}u'\right)^2=0.
\label{eqKSsubeqgenus0last}
\end{eqnarray}

\`A l'\'etape 4,
on trouve alors la valeur un pour le genre de (\ref{eqKSsubeqgenus1}), 
et la valeur z\'ero pour celui de 
(\ref{eqKSsubeqgenus0first})--(\ref{eqKSsubeqgenus0last}),
puis les expressions (\ref{eqKSElliptic}) et (\ref{eqKSTrigo})
pour les solutions.

Avec la valeur minimale $m=3$,
on retrouve donc tous les r\'esultats connus.
Le calcul pour $m \ge 4$ est actuellement en cours.

\section{Conclusion}

L'\'etude par approximants de Pad\'e semble conforter la nature 
probablement elliptique de l'unique solution analytique de codimension
nulle de l'EDO chaotique (\ref{eqKSODE}).
Parmi les autres pistes, mi-analytiques, mi-num\'eriques,
qu'il serait possible de mettre en {\oe}uvre,
il en existe une particuli\`erement int\'eressante car constructive,
c'est celle d'obtention de la sous-\'equation par la s\'erie de Laurent.
Son seul inconv\'enient est l'absence de borne sup\'erieure pour le degr\'e
$m$,
ce qui contraint \`a de volumineux calculs qui sont encore en cours.

\section*{Remerciements}

Nous remercions les organisateurs et l'IRMA pour l'invitation \`a donner 
cette conf\'eren\-ce,
Karim Belabas pour plusieurs suggestions fructueuses,
ainsi que 
Jean-Robert Burgan, 
Chiang Yik-man, 
Pierre Moussa, 
Jean-Pierre Ramis,
Simon Ruijnesaars 
 et 
Jean Thomann
pour des discussions enrichissantes.
La facilit\'e de programmation du langage de calcul formel 
AMP \cite{Drouffe} a constitu\'e une aide pr\'ecieuse.

T.-L.~Yee remercie la Fondation Croucher de Hong-Kong,
et MM le CEA pour leur soutien financier.

\vfill \eject


\vfill \eject

\begin{thebibliography}{99}   

\bibitem{AgrawalBook} G.~P.~Agrawal,
\textit{Nonlinear fiber optics}, 3rd edition
(Academic press, Boston, 2001).

\bibitem{PadeIntro} C.~Brezinski, J.~van Isegehm,
Pad\'e Approximations,
47--222,
\textit{Handbook of Numerical Analysis},
Vol.~III", 
eds.~P.G.~Ciarlet and J.-L.~Lions (North-Holland, Amsterdam, 1994).

\bibitem{Cargese1996} R.~Conte (ed.),
\textit{The Painlev\'e property, one century later}, 
810 pages,
CRM series in mathematical physics (Springer, New York, 1999).

\bibitem{CFP1993} R.~Conte, A.~P.~Fordy, and A.~Pickering,
A perturbative Painlev\'e approach to nonlinear differential equations,
Physica D {\bf 69} (1993) 33--58.

\bibitem{CM1989} R.~Conte and M.~Musette,
Painlev\'e analysis and B\"acklund transformation in the 
Kuramoto--Sivashinsky equation,
J.~Phys.~A {\bf 22} (1989) 169--177.

\bibitem{Drouffe} J.-M.~Drouffe,
AMP reference manual, version 9 (1993),
SPhT, CE Saclay, F-91191 Gif-sur-Yvette Cedex.
 
\bibitem{Eremenko1986} A.~E.~Eremenko, 
Meromorphic solutions of equations of Briot-Bouquet type,
Teor.~Funktsii, 
Funktsional'nyi Analiz i Prilozhen.~Vyp.~{\bf 16} (1982) 48--56
    [English~: Amer.~Math.~Soc.~Transl.~{\bf 133} (1986) 15--23].

\bibitem{FournierSpiegelThual} 
J.-D.~Fournier, E.~A.~Spiegel, and O.~Thual,  
Meromorphic integrals of two nonintegrable systems,
{\it Nonlinear dynamics},
366--373,
ed.~G.~Turchetti (World Scientific, Singapore, 1989).

\bibitem{GP1997} J.~Gilewicz and M.~Pindor,
Pad\'e approximants and noise: a case of geometric series,
J.~Comput.~Appl.~Math. {\bf 87} (1997) 199--214.

\bibitem{GP1999} J.~Gilewicz and M.~Pindor,
Pad\'e approximants and noise: rational functions,
J.~Comput.~Appl.~Math. {\bf 105} (1999) 285--297.

\bibitem{vHSvS} M.~van Hecke, C.~Storm, and W.~van Saarlos,         
Sources, sinks and wavenumber selection in coupled CGL equations and
experimental implications for counter-propagating wave systems,
Physica D {\bf 133} (1999) 1--47. Patt--sol/9902005.

\bibitem{MapleAlgcurves} Mark van Hoeij,
progiciel ``algcurves'', Maple V (1997).
\verb+http://www.math.fsu.edu/~hoeij/algcurves.html +

\bibitem{js1998} A.~Joux and J.~Stern, 
Lattice reduction: a toolbox for the cryptanalyst,
J.~Cryptology {\bf 11} (1998) 161--185.

\bibitem{Knuth} D.~Knuth,
The art of computer programming, 
3rd ed.~ (Addison-Wesley, Reading (Mass.), 1998).

\bibitem{KudryashovKSFourb} N.~A.~Kudryashov,
Exact soliton solutions of the generalized evolution equation 
of wave dynamics,
Prikladnaia Matematika i Mekhanika {\bf 52} (1988) 465--470
[English~: 
Journal of applied mathematics and mechanics {\bf 52} (1988) 361--365].

\bibitem{Kud1990} N.~A.~Kudryashov,                      
Exact solutions of the generalized Kuramoto--Sivashinsky equation,
Phys.~Lett.~A {\bf 147} (1990) 287--291.

\bibitem{KuramotoTsuzuki} Y.~Kuramoto and T.~Tsuzuki,
Persistent propagation of concentration waves in dissipative media far from
thermal equilibrium,
\PTP\ {\bf 55} (1976) 356--369.

\bibitem{LegaThese} J.~Lega, Th\`ese (Universit\'e de Nice, 28 mars 1989).

\bibitem{Lega2001} J.~Lega,                                    
Traveling hole solutions of the complex Ginzburg-Landau equation: a review,
Physica D {\bf 152--153} (2001) 269--287.

\bibitem{lll1982} A.~K.~Lenstra, H.~W.~Lenstra, Jr., and Lov\'asz,
Factoring polynomials with rational coefficients,
Math.~Ann.~{\bf 261}  (1982) 515--534. 

\bibitem{MannevilleBook} P.~Manneville,
\textit{Dissipative structures and weak turbulence}
(Academic Press, Boston, 1990).
Enlarged French translation:
\textit{Structures dissipatives, chaos et turbulence}
(Al\'ea-Saclay, Gif-sur-Yvette, 1991).

\bibitem{Maple} Maple, 
\verb+http://www.maplesoft.com/products/Maple8/index.shtml +

\bibitem{MC2003} M.~Musette and R.~Conte,
Analytic solitary waves of nonintegrable equations,
9 pages,
Physica D {\bf } (2003), accept\'e pour publication.
S2002/069.
nlin.PS/0302051

\bibitem{PaiLecons} P.~Painlev\'e,                               
{\it Le\c{c}ons sur la th\'eorie analytique des \'equations diff\'erentielles}
(Le\c{c}ons de Stockholm, 1895)
(Hermann, Paris, 1897).
R\'eimpression, {\it O$\!$euvres de Paul Painlev\'e}, vol.~I
(\'Editions du CNRS, Paris, 1973).
 
\bibitem{PM1979} Y.~Pomeau and P.~Manneville,                    
Stability and fluctuations of a spatially periodic flow,
J.~Physique Lett.~{\bf 40} (1979) L609--L612.

\bibitem{Porubov1996} A.~V.~Porubov,                                
Periodical solution to the nonlinear dissipative equation for surface waves
in a convecting liquid layes,
Phys.~Lett.~A {\bf 211} (1996) 391--394.

\bibitem{NumericalRecipes} W.~H.~Press, S.~A.~Teukolsky, W.~T.~Vetterling, 
and B.~P.~Flannery,
\textit{Numerical recipes in C}, second edition
(Cambridge university press, Cambridge, 1992).

\bibitem{Ruelle} D.~Ruelle,
\textit{Hasard et chaos} (Odile Jacob, Paris, 1991).

\bibitem{Samsonov1994} A.~M.~Samsonov,                               
Nonlinear strain waves in elastic waveguides,
{\it Nonlinear waves in solids}, 349--382
eds.~A.~Jeffrey and J.~Engelbrecht
(Springer-Verlag, Wien, 1994). 

\bibitem{TY} J.~Thomann et J.-C.~Yakoubsohn,
communication priv\'ee (2002). 

\bibitem{TF} O.~Thual and U.~Frisch,                                 
Natural boundary in the Kuramoto model,
{\it Combustion and nonlinear phenomena},
327--336,
eds.~P.~Clavin, B.~Larrouturou, and P.~Pelc\'e
(\'Editions de physique, Les Ulis, 1986).

\bibitem{Valiron} G.~Valiron,                                        
{\it Cours d'analyse math\'ematique},
2i\`eme \'ed. (Masson, Paris, 1950).

\end{thebibliography}
\end{document}